\documentclass[sigconf]{acmart}

\AtBeginDocument{%
  }

\copyrightyear{2025}
\acmYear{2025}
\setcopyright{acmlicensed}
\acmConference[DHOW '25] {Proceedings of the 2nd International Workshop on Diffusion of Harmful Content on Online Web}{October 27--31, 2025}{Dublin, Ireland.}
\acmBooktitle{Proceedings of the 2nd International Workshop on Diffusion of Harmful Content on Online Web (DHOW '25), October 27--31, 2025, Dublin, Ireland}
\acmISBN{979-8-4007-2057-4/2025/10}
\acmDOI{10.1145/XXXXXX.XXXXXX}

\usepackage{paralist}
\usepackage{wrapfig}
\usepackage{hyperref}

\begin{document}

\title[Societal Impact of Deepfakes in Low-Tech Environments]{Seeing Isn't Believing: Addressing the Societal Impact of Deepfakes in Low-Tech Environments}

\author{Azmine Toushik Wasi}
\orcid{0000-0001-9509-5804}
\authornote{Equal Contribution.}
\authornote{Corresponding author.}
\affiliation{%
  \institution{Shahjalal University of Science and Technology}
  \city{Sylhet}
  \country{Bangladesh}}
  \email{azmine32@student.sust.edu}

\author{Rahatun Nesa Priti}
\orcid{0009-0002-6441-4512}
\authornotemark[1]
\affiliation{%
  \institution{Shahjalal University of Science and Technology}
  \city{Sylhet}
  \country{Bangladesh}}
  \email{rahatunnesa2092@gmail.com}

\author{Mahir Absar Khan}
\orcid{0009-0004-8147-209X}
\authornotemark[1]
\affiliation{%
  \institution{Shahjalal University of Science and Technology}
  \city{Sylhet}
  \country{Bangladesh}}
  \email{mahir52@student.sust.edu}

\author{Abdur Rahman}
\orcid{0009-0002-0507-8870}
\authornotemark[1]
\affiliation{%
  \institution{Shahjalal University of Science and Technology}
  \city{Sylhet}
  \country{Bangladesh}}
  \email{abdur37@student.sust.edu}

\author{Mst Rafia Islam}
\orcid{0009-0000-2561-3418}
\affiliation{%
  \institution{Independent University}
  \city{Dhaka}
  \country{Bangladesh}}
\email{rafiabarsha21@gmail.com}

\renewcommand{\shortauthors}{Wasi et al.}

\begin{abstract}
Deepfakes, AI-generated multimedia content that mimics real media, are becoming increasingly prevalent, posing significant risks to political stability, social trust, and economic well-being, especially in developing societies with limited media literacy and technological infrastructure. This work aims to understand how these technologies are perceived and impact resource-limited communities. We conducted a survey to assess public awareness, perceptions, and experiences with deepfakes, leading to the development of a comprehensive framework for prevention, detection, and mitigation in tech-limited environments. Our findings reveal critical knowledge gaps and a lack of effective detection tools, emphasizing the need for targeted education and accessible verification solutions. This work offers actionable insights to support vulnerable populations and calls for further interdisciplinary efforts to tackle deepfake challenges globally, particularly in the Global South.
\end{abstract}

\ccsdesc[500]{Human-centered computing~Collaborative content creation}
\ccsdesc[500]{Human-centered computing~Social media}
\ccsdesc[300]{Human-centered computing~Social content sharing}
\ccsdesc[500]{Human-centered computing~HCI design and evaluation methods}
\ccsdesc[300]{Applied computing~Law, social and behavioral sciences}

\keywords{Deepfakes, Global South, Media Literacy, AI Detection Tools, Misinformation}
%
%
%


\maketitle

\section{Introduction}
Deepfakes, a form of AI-generated multimedia content, encompass images, videos, audio, and text designed to convincingly mimic real media \cite{17}. The term combines "deep learning," an advanced artificial intelligence technique, with "fake," reflecting how sophisticated algorithms are leveraged to manipulate or generate content \cite{7}. This technology is highly proficient in producing realistic material, often blurring the line between reality and fabrication \cite{Dunnell2024-zu}. While deepfakes have legitimate applications in fields such as entertainment, education, and advertising \cite{16}, their misuse has triggered widespread societal concerns. Instances of identity theft, fraud, revenge tactics, and national security threats highlight the darker implications of this technology \cite{Lee2024-qv}.
Deepfake generation relies on AI models trained on extensive datasets to produce highly realistic outputs. A common example is deepfake videos that depict individuals saying or doing things they never actually did. On the positive side, this technology facilitates innovations such as virtual try-ons in fashion and realistic special effects in the film industry \cite{1} \cite{Clarke2023-ds} \cite{Danry2022-el}. However, its potential for misuse is equally significant, with deepfakes capable of deceiving audiences and causing personal and societal harm. This duality underscores the urgent need to assess both the opportunities and risks associated with deepfakes, as well as to develop effective mitigation strategies \cite{Gamage2022-jb}.

\begin{figure}
  \centering
    
    \includegraphics[scale=0.35]{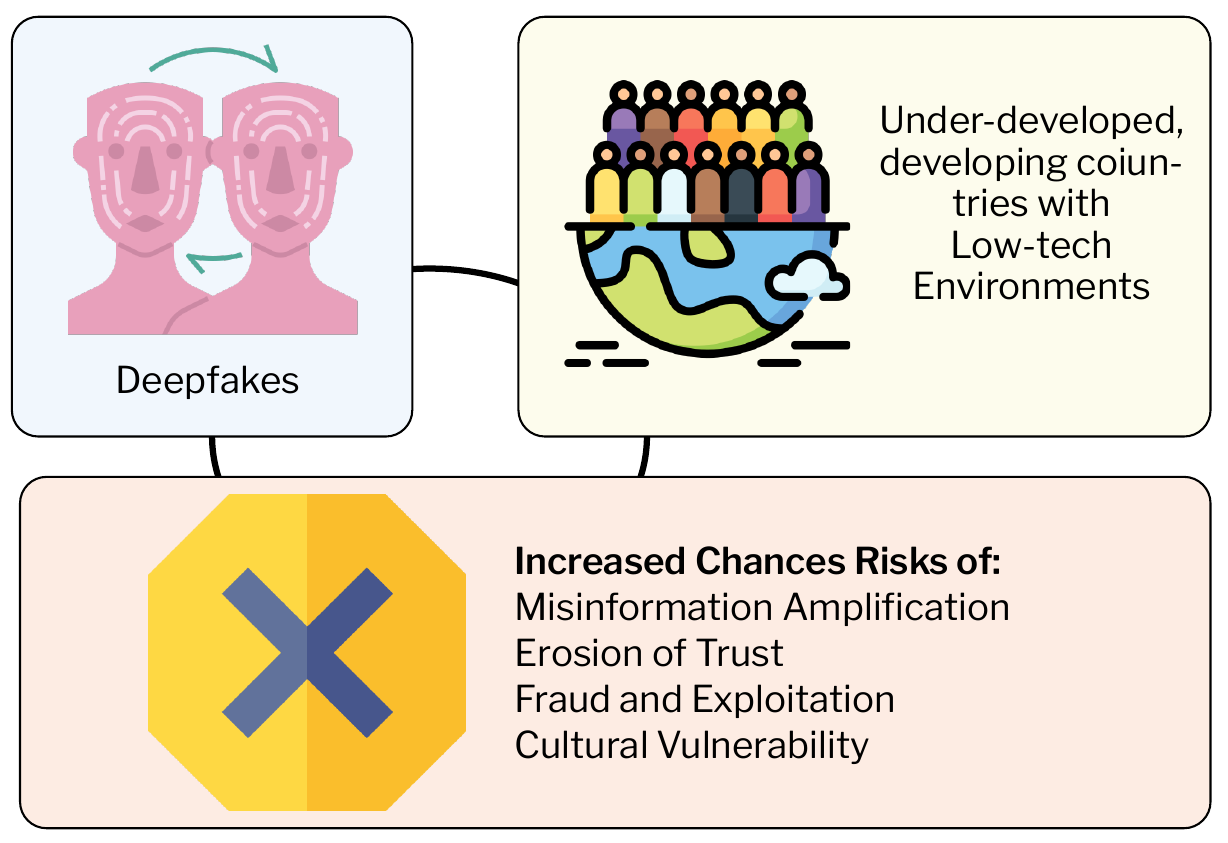}
    
    \caption{Motivation behind this study is to understand the adverse impact of deepfakes in tech-limited environments, where limited access to digital resources and media literacy exacerbate the risks posed by AI-generated content.}
   \label{fig:WHAT}
    
\end{figure}

The rise of deepfake technology has raised substantial concerns, particularly in underdeveloped and developing countries with low-tech environments \cite{ALKHAZRAJI2023}. Deepfakes, hyper-realistic AI-generated media, pose significant risks to political stability, social trust, and economic conditions in regions with low media literacy and limited capacity to counteract such threats. A key concern is the dissemination of misinformation and the manipulation of public opinion, especially in politically fragile nations where deepfakes can create false narratives that undermine trust in leadership and influence electoral outcomes \cite{Qureshi2024}. For instance, fabricated videos of public figures making controversial statements can lead to public unrest and disrupt democratic processes. The proliferation of deepfake content through social media exacerbates these challenges, particularly in areas with limited access to reliable news sources, further eroding trust in governance and traditional media \cite{Christofoletti2024}.

Beyond political manipulation, deepfakes contribute to a broader crisis of media credibility. As awareness grows about the ease of altering digital content, public skepticism increases, especially in democratic societies reliant on informed citizens \cite{Sohrawardi2024}. In regions with limited media literacy and verification resources, individuals are more vulnerable to deception, heightening societal divisions and confusion during critical events such as elections and protests \cite{YuriiDudka2023}.
Economically, deepfakes introduce significant risks, particularly in fragile economies where fraudulent activities such as identity theft and financial scams can lead to severe financial losses \cite{Hummer2023}. Developing nations often lack the resources to combat these threats effectively through legal measures, awareness campaigns, and detection technologies.
Culturally, deepfakes have the potential to deepen existing societal tensions by targeting specific ethnic, religious, or political groups, thereby exacerbating social discord and undermining communal harmony \cite{Achyut2023}.
The urgency of addressing deepfake-related challenges is amplified by the rapid digital expansion in developing countries, where the growing penetration of the internet and social media lacks parallel advancements in regulatory frameworks and technological defenses. Without adequate countermeasures, these regions remain highly vulnerable to the harmful consequences of deepfakes, which threaten social cohesion, political stability, and economic resilience.

In this paper, we present a comprehensive framework to address the societal impacts \cite{Gamage2022-dw} and cross-disciplinary vulnerabilities of deepfake technology, with a specific focus on tech-limited environments. Our contributions include conducting an in-depth survey to explore individual and community perceptions of deepfakes, uncovering concerns related to awareness, coping mechanisms, and the broader societal impact. Based on these insights, we propose a three-stage framework, \textbf{Prevention, Detection, and Mitigation}, that provides an end-to-end solution to combat deepfake challenges. The \textbf{Prevention} stage focuses on ethical considerations, regulatory measures, and public awareness campaigns to reduce harmful deepfake creation. The {Detection} stage emphasizes AI-driven tools and community-based initiatives for identifying manipulated content in real time. Lastly, the \textbf{Mitigation} stage outlines strategies to manage the aftermath of deepfake dissemination through legal responses, educational programs, and ethical guidelines. Our work offers practical, context-aware solutions to empower communities, policymakers, and stakeholders in effectively countering deepfake threats in resource-constrained settings.

\section{Methodology}  
Our methodology consists of three key steps: conducting a survey to understand perceptions of deepfakes in low-tech environments, collecting data on their societal impact, and developing a framework to address these challenges with practical solutions.  

\subsection{Data Collection}
\noindent
\textbf{Survey Design.}  
The survey aimed to assess awareness, perceptions, and societal impacts of deepfakes in a developing society with limited technological infrastructure. It explored familiarity with deepfakes, frequency of exposure, and observed or experienced consequences. Additionally, the survey examined concerns about the potential misuse of deepfakes in misleading and manipulating public opinion. The findings aimed to highlight areas requiring educational initiatives, policy interventions, and countermeasures to mitigate deepfake-related challenges.  

\noindent
\textbf{Data Collection and Participant Demographics.}
We collected 73 responses from participants in Bangladesh, primarily from urban areas (75.3\%), with 23.3\% from suburban and 1.4\% from rural regions, representing a technologically exposed demographic. The gender distribution was nearly balanced (52.1\% female, 47.9\% male), and the majority (83.6\%) were aged 18-24, with 16.4\% in the 25-34 range, focusing on younger, digitally active individuals. Most respondents (98.6\%) had higher education, and 90.4\% were students, reflecting a highly educated, tech-savvy group.

\begin{figure*}
    \centering
    
    \includegraphics[width=\linewidth]{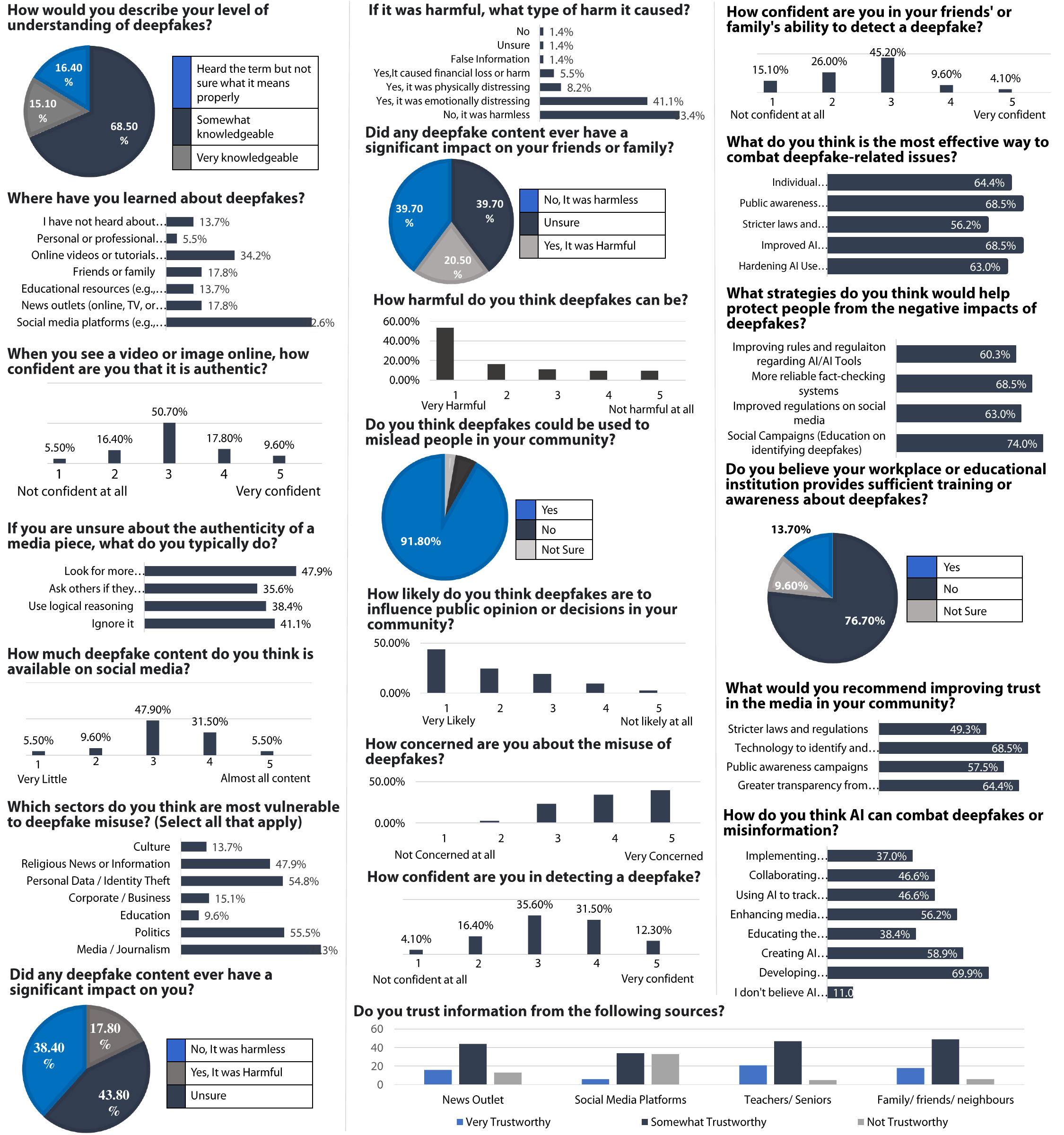}
    
    \caption{Data Analysis: Statistics.}
    \label{fig:Stat}
    
\end{figure*}

\subsection{Framework Design}
After designing the survey and collecting data, we analyze the findings to develop a framework that addresses concerns, incorporates suggestions, and aligns with regulatory changes.

\subsubsection{Understanding User Perceptions.}
Deepfakes significantly impact social dynamics in developing societies, particularly where fact-checking resources are limited. These hyperrealistic media can manipulate public perception, incite unrest, and threaten political stability and information integrity, especially in fragile democracies \cite{Noor2024}. The rapid spread of misinformation through social media exacerbates the issue, with deepfakes often targeting political figures to undermine their credibility \cite{Veerasamy2022,2023}\cite{Tahir2021-yv}. For instance, during Nigeria's recent elections, deepfakes were used for mudslinging and propaganda, distorting political narratives \cite{2023}. As deepfake technology becomes more sophisticated, concerns over trust and social interaction rise, as individuals struggle to differentiate reality from manipulation \cite{DoanAkkaya2024,Fehring2023}.

Survey responses reveal growing awareness of deepfakes, particularly in tech-limited societies. 68.5\% of respondents were somewhat familiar with deepfakes, and 15.1\% were very knowledgeable, but a significant knowledge gap remains, with 16.4\% hearing the term but lacking understanding. This highlights the need for public education on deepfakes and their consequences. Notably, 72.1\% had encountered deepfake content, even in low-tech environments. Alarmingly, 17.8\% experienced harm, while 43.8\% were unsure of its impact, suggesting harm may be underreported due to lack of awareness. Emotional distress (41.1\%) was the most common consequence, followed by financial (5.5\%) and physical distress (8.2
Societal ripple effects are significant: 20.5\% believed deepfakes had impacted their friends or family, while 39.7\% were unsure, indicating potential underestimation of indirect harm. A large majority (91.8\%) expressed concern over deepfakes misleading their community. Additionally, 53.4\% perceived deepfakes as harmful, and 73.9\% were worried about their misuse. Notably, 43.8\% believed deepfakes could influence public opinion and decisions, linking exposure to deepfakes with perceived societal consequences. These findings underscore the urgent need for research and intervention in tech-limited societies, where misinformation can spread unchecked and countermeasures are scarce. A cross-disciplinary approach involving technology, sociology, psychology, and policy-making is essential to mitigate deepfake risks and protect vulnerable populations. Detailed statistics are available in Figure \ref{fig:Stat}.

\subsubsection{Connecting the Dots.}
In this section, we connect the survey findings to actionable insights, bridging perceptions with components to consider for the framework.

\noindent
\textbf{Low Confidence in Deepfake Detection.}\quad
The survey reveals a significant gap in respondents' ability to detect deepfakes\cite{Shahid2022-qv}, with 16.4\% expressing low confidence and 4.1\% reporting no confidence at all. This uncertainty extends to the respondents' perceptions of their acquaintances' detection skills, with 45.2\% expressing moderate confidence and 26\% indicating low confidence. These findings highlight a critical need for awareness and training in deepfake detection. Public education campaigns should be launched to equip individuals with the tools and techniques needed to identify AI-generated content\cite{Sohrawardi2024}. Furthermore, educational programs could be incorporated into schools, workplaces, and public spaces to promote digital literacy and content verification. By addressing this knowledge gap, communities will be better prepared to protect themselves from the risks posed by deepfakes.

\noindent
\textbf{Gaps in Information Verification Practices.}\quad 
Although 69.9\% of respondents report cross-checking information with other sources, 21.9\% do not engage in verification at all. This indicates a lack of awareness or resources for verifying content, which could be mitigated by providing accessible, AI-powered verification tools. These tools could be integrated into social media platforms, news websites, and digital services, enabling real-time content verification. Additionally, educational programs focusing on the importance of information verification should be prioritized to help individuals understand the significance of skepticism and develop the habit of cross-checking content before sharing. Ensuring that people are equipped with these practices is vital in a world where misinformation spreads rapidly.

\noindent
\textbf{Trust Issues in Media and Misinformation.}\quad 
The survey highlights widespread concern about the use of deepfakes to manipulate public perception, with 91.8\% of respondents expressing concern about their potential to deceive. This reflects broader mistrust in media and digital content, exacerbated by deepfake technology. To rebuild trust, media outlets must prioritize transparency by clearly labeling AI-generated or manipulated content. This transparency would allow audiences to assess the authenticity of the media they consume. Public awareness campaigns should also be implemented to teach people how to critically evaluate digital content and recognize potential manipulation. By fostering a culture of responsible media consumption, these campaigns can help reduce misinformation and restore trust in media organizations.

\noindent
\textbf{Lack of Confidence in AI Detection Tools.}\quad
While 50\% of respondents support the use of AI detection tools, concerns about their effectiveness persist. To address these doubts, it is essential to make AI detection tools widely accessible and user-friendly. These tools should be integrated into popular platforms such as social media, video-sharing sites, and news outlets to facilitate real-time content verification. Public training sessions or online tutorials could also be provided to ensure individuals know how to use these tools effectively\cite{Usmani2024-os}. Ongoing development and updates to these tools will be crucial in keeping pace with evolving deepfake technologies. By ensuring accessibility and reliability, these tools can boost public confidence in detecting and mitigating deepfake-related misinformation.

\noindent
\textbf{Concerns About Regulation and Accountability.}\quad
While 63\% of respondents support better regulation of deepfakes, there is less enthusiasm for regulating AI technologies themselves. This suggests that while there is recognition of the need for oversight, people may be wary of overregulation that could stifle technological innovation. A balanced approach to regulation is necessary, one that addresses malicious uses of deepfake technology without impeding the progress of beneficial AI advancements. Governments, tech companies, and civil society organizations should collaborate to create clear, targeted regulations that safeguard public safety and ethical standards while promoting innovation. Moreover, regulatory bodies should work with AI developers to ensure that detection tools are regularly updated to counter emerging threats from deepfake content.

\noindent
\textbf{Strategies to Restore Trust in Media.}\quad
The survey indicates that respondents see AI-powered deepfake detection as the most effective strategy for restoring trust in media, with 68.5\% supporting its implementation. Transparency from media outlets, improved regulations, and public awareness campaigns also received significant backing. To restore trust, AI-driven fact-checking systems should be deployed across digital platforms to flag deepfakes and misinformation in real-time. Media organizations must also adopt transparent policies, clearly marking AI-generated or altered content to help viewers distinguish between authentic and manipulated media. Combined with public education on media literacy and the importance of verifying information, these strategies will work together to mitigate the effects of deepfakes and rebuild trust in the media.

\begin{figure*}
    \centering
    \includegraphics[width=\linewidth]{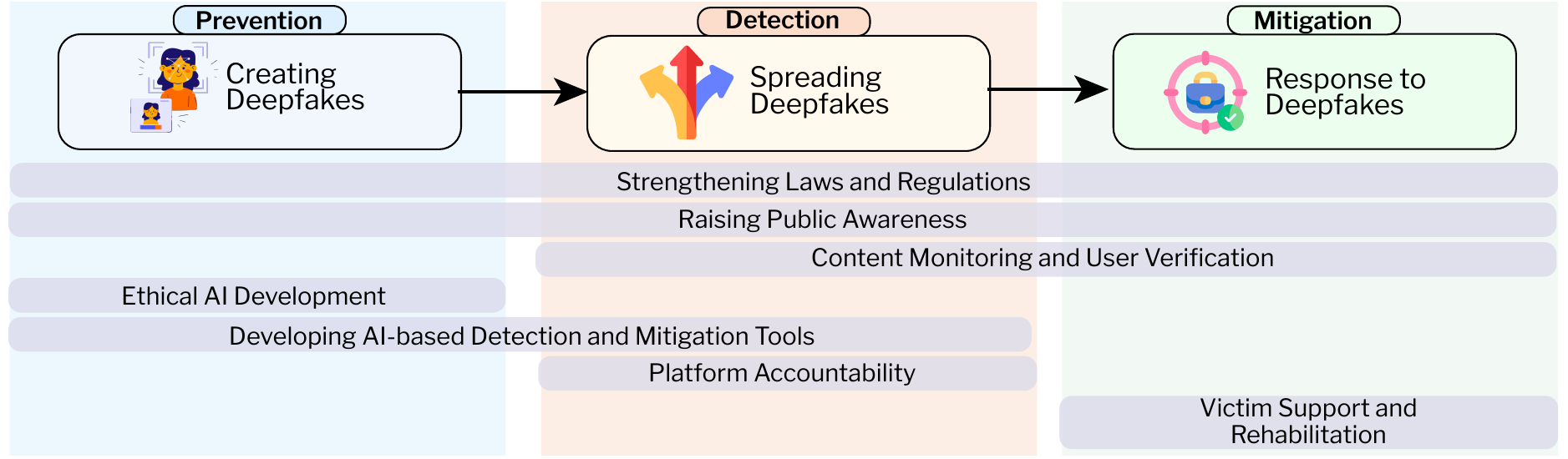}
    
    \caption{Our framework for handling deepfakes in low-tech environments consists of three key stages: prevention, detection, and mitigation. In the prevention stage, we focus on the ethical challenges and technological factors involved in the creation of deepfakes, aiming to reduce their emergence and impact.}
    \label{fig:Framework}
    
\end{figure*}

\section{Framework} 
This section presents the components and stages of the framework designed to address the challenges posed by deepfakes. It is divided into three key stages: Creating Deepfakes, Spreading Deepfakes, and Responding to Deepfakes, each discussed with associated components and strategies to manage risks effectively.

\subsection{Prevention: Preventing from \textit{Creating Deepfakes}} 
This stage focuses on the components related to deepfake creation, including strategies to control deepfake generation and methods to raise public awareness to avoid misuse.

\subsubsection{Strengthening Laws and Regulations} 
Laws regulating deepfakes remain in development and vary across regions. The European Union's AI Act, for example, introduces transparency obligations for creators, requiring disclosure when content is AI-generated or manipulated \cite{deAlmeida2021}. However, it addresses use more than creation. In the U.S., tort laws, such as "appropriation of name or likeness" and "false light publicity," can address harm caused by deepfakes, but challenges remain, including proving harm and balancing with First Amendment rights \cite{Mania2022}. States like Indiana and Florida have passed laws targeting fabricated media and deepfakes involving minors \cite{Mania2022}. Australia has introduced legislation against non-consensual deepfake sexual content. A global framework regulating both creation and distribution is needed, combining privacy, intellectual property, and ethical considerations. Strengthening the alignment of existing laws and AI-specific legislation is crucial to address these multifaceted challenges, such as copyright and privacy protections \cite{Widder2022}. Clear legal frameworks, enforcement, and international cooperation are key to mitigating the misuse of deepfakes.

\subsubsection{Raising Public Awareness.} Raising public awareness about deepfakes is essential to preventing misuse and mitigating harm. Educational campaigns should inform the public about the risks of deepfakes, such as disinformation and privacy violations \cite{Tan2024}. Collaboration between platforms, policymakers, and media organizations can help promote ethical guidelines and detection technologies, empowering users to critically evaluate content \cite{Jiang2024}. Legal frameworks like the Deepfake Accountability Act could also raise awareness by holding creators accountable and promoting ethical practices \cite{Masood2022}.

\subsubsection{Ethical AI Development.} A multidimensional approach to ethical AI development is essential to prevent harmful deepfakes. Developers should establish clear guidelines prohibiting malicious uses, such as non-consensual pornography and misinformation campaigns \cite{sathaye2024deepfake}. This involves implementing safeguards, robust access controls, and auditing mechanisms for AI systems \cite{Masood2022}. Collaboration with ethicists, legal experts, and affected communities ensures alignment with societal norms and values. While transparency is important for accountability, it must be paired with ethical constraints to prevent exploitation \cite{Li2021}. Techniques like watermarking and automated detection tools can further deter misuse \cite{Verdoliva2020}. Promoting ethical AI, through public awareness and regulatory oversight, prioritizes societal well-being and combats digital deception \cite{Widder2022}.

\subsubsection{Developing AI-based Detection and Mitigation Tools.} AI-based detection and mitigation tools are crucial for preventing deepfake creation by identifying manipulations during content creation and distribution. Convolutional and recurrent neural networks can detect subtle discrepancies in videos, such as unnatural facial expressions, lighting inconsistencies, or pixel-level changes \cite{Jiang2024}\cite{Han2024-oi}. Integrating these tools into content-sharing platforms helps prevent malicious content from going viral \cite{Tan2024}. Monitoring physiological cues, such as eye blinking and lip-syncing inconsistencies, adds another layer of protection \cite{Dolhansky7}. AI models can also analyze metadata to detect post-creation anomalies, ensuring content authenticity \cite{Whyte2020}. Watermarking technologies, including blockchain solutions, help trace media origins and complicate manipulation \cite{Solaiman2024}. Compression artifact detection tools, such as GANalyzer, expose flaws in deepfake-generated content \cite{Akhtar2024}.

\subsection{Detection: Detecting and Stopping the \textit{Spread of Deepfakes}}
This stage explores measures to prevent the spread of deepfakes once created, including real-time detection technologies, content removal systems, and platform collaborations to enforce these measures.

\subsubsection{Strengthening Laws and Regulations.} Current laws addressing deepfakes remain fragmented. U.S. laws like \href{https://statutes.capitol.texas.gov/Docs/EL/htm/EL.274.htm}{Texas' amended Election Code} and \href{https://leginfo.legislature.ca.gov/faces/billNavClient.xhtml? bill_id=202320240AB2839}{California's Deceptive Audio or Visual Media Act} criminalize deepfakes aimed at deceiving voters \cite{Farish2019}. The EU's proposed AI Act mandates transparency, requiring disclosure of AI manipulation \cite{deAlmeida2021}, but enforcement and prevention remain challenges. Privacy laws such as the GDPR protect likenesses but don't fully address deepfake proliferation \cite{Hoofnagle2019}. A holistic approach integrating content moderation, stricter penalties, and real-time detection systems is necessary to ensure rapid removal and global enforcement.

\subsubsection{Raising Public Awareness.} 

To address the risks of deepfakes, enhancing public awareness before their spread is vital. A comprehensive strategy includes promoting digital literacy, ethical guidelines, and educational initiatives to help the public recognize manipulated content. Social media platforms, such as Instagram with its educational campaigns and reporting features, demonstrate how proactive measures can inform users \cite{Cochran2021}. Integrating media literacy into curricula, like Google's Digital Literacy and Citizenship program, enhances critical thinking and helps students evaluate online content \cite{Whyte2020}. Schools can further support this through workshops and seminars on identifying fake media. Accessible detection tools, such as Deepware Scanner and Sensity AI, empower users to verify media authenticity \cite{Wang2024}. Public awareness campaigns, such as the EU's \texttt{StopFakeNews} initiative, educate wider audiences about deepfake risks through diverse media channels \cite{SamuelOkon2024}. Platforms must also invest in AI-powered detection systems, collaborating with initiatives like the Deepfake Detection Challenge to flag manipulated content in real-time \cite{Wang2024}. Governments can contribute by enacting laws like the Deepfake Accountability Act and the EU's Digital Services Act to hold platforms accountable and mandate pre-upload AI checks for deepfakes \cite{Whittaker2023}. Combining education, technology, legislation, and proactive platform policies creates a multifaceted approach to reducing deepfake creation and spread, fostering a more informed society.

\subsubsection{Content Monitoring and User Verification.} 

Content monitoring and user verification are crucial measures for preventing the spread of deepfakes on platforms like YouTube, Facebook, and Twitch. By utilizing advanced detection technologies, these platforms can identify and remove harmful content before it reaches a large audience. For example, GAN Fingerprinting helps identify deepfakes by analyzing patterns unique to Generative Adversarial Networks (GANs), allowing platforms to trace the source of manipulations \cite{frank2021}. Furthermore, AI-based models such as Convolutional Neural Networks (CNNs) and Recurrent Neural Networks (RNNs) analyze video frames to detect inconsistencies in facial expressions or skin texture \cite{Jiang2024}. Physiological analysis tools like DeepFake-o-meter\footnote{\hyperlink{https://zinc.cse.buffalo.edu/ubmdfl/deep-o-meter}{zinc.cse.buffalo.edu/ubmdfl/deep-o-meter}} can also spot unnatural eye-blinking patterns \cite{Dolhansky7}, while metadata analysis can detect discrepancies in timestamps or camera details, helping verify the authenticity of media \cite{Solaiman2024}. Platforms can integrate watermarking techniques, including blockchain-based digital watermarks, to ensure content integrity \cite{Whyte2020}. Audio-visual synchronization tools, like Microsoft's Video Authenticator, help detect mismatches between speech and lip movements \cite{Akhtar2024}. By combining these technologies with user verification measures, such as account authentication and behavior monitoring, platforms can create a safer digital environment, reduce the likelihood of deepfake spread, and uphold legal and ethical standards in content distribution. Fact-checkers like PolitiFact\footnote{\hyperlink{https://www.politifact.com}{politifact.com}} and FactCheck.org\footnote{\hyperlink{https://www.factcheck.org}{factcheck.org}} are essential in combating deepfakes by using expert analysis and AI tools to verify content authenticity. They cross-reference manipulated material with trusted sources, such as public appearances or reputable news outlets, to identify inconsistencies. For example, when a deepfake portrays a politician making false statements, fact-checkers compare it with existing transcripts and reports to expose discrepancies. Platforms like Facebook collaborate with these organizations to label deepfake content, offering context to users. Additionally, crowd-sourcing through user reports helps platforms like YouTube and Twitter quickly verify suspicious content with input from experts and AI systems, improving the fight against deepfakes \cite{Masood2022,YadlinSegal2020,Hoofnagle2019,Widder2022}.

\subsubsection{Developing AI-based Detection and Mitigation Tools.} AI-based detection and mitigation tools are essential in preventing the spread of deepfakes, especially on platforms like YouTube, Facebook, and Twitch. By leveraging machine learning, AI can identify deepfake content before it gains widespread attention. GAN Fingerprinting detects patterns unique to Generative Adversarial Networks (GANs), such as pixel-level inconsistencies, enabling early identification and source tracing \cite{frank2021}. Blockchain technology secures digital content by embedding cryptographic signatures to verify authenticity and ensure media integrity post-distribution \cite{Whyte2020}. AI systems like Microsoft's Video Authenticator use neural networks to analyze facial and motion inconsistencies, assigning confidence scores to manipulation likelihood \cite{Heidari2024}. Multi-modal detection systems analyze both visual and audio components, improving detection accuracy by cross-verifying inconsistencies \cite{KiChan2020}. Real-time detection tools, such as TrueMedia, monitor live-streamed content for deepfake signs, crucial during events like elections \cite{Tan2024}. Audio forensic analysis tools identify synthetic audio manipulations, offering a comprehensive approach to media verification. These technologies help detect and prevent deepfakes from spreading, contributing to a safer digital environment as they evolve.

\subsubsection{Platform Accountability.} Ensuring platform accountability for managing deepfakes involves a combination of legislative actions, technical measures, and corporate responsibility. Laws like the U.S. Deepfake Accountability Act\footnote{https://www.congress.gov/bill/118th-congress/house-bill/5586/text} and the EU's Digital Services Act\footnote{https://www.eu-digital-services-act.com/} hold platforms accountable for hosting harmful content, while transparency initiatives, such as labeling AI-generated content, enhance trust. Major tech companies, including Meta and IBM, are investing in AI tools for real-time deepfake detection and collaboration with cybersecurity experts. Platforms use advanced detection methods like GAN fingerprinting and multi-modal systems to identify manipulations, while combining automated systems with human oversight to prevent the spread of harmful content \cite{Tan2024} \cite{Zhang2022}.

\subsection{Mitigation: \textit{Response to Deepfakes}}
This stage addresses strategies for responding to deepfakes after their spread, focusing on detection tools, debunking false media, and issuing corrections through trusted channels.

\subsubsection{Strengthening Laws and Regulations.}
Existing laws like U.S. tort laws (e.g., "false light publicity") provide avenues for victims of deepfakes to seek redress \cite{Mania2022}. In the EU, the AI Act's transparency requirements could help curb deepfake spread by making creators disclose AI manipulation \cite{deAlmeida2021}. GDPR can also aid in removing manipulated content that violates personal data protections \cite{Hoofnagle2019}. However, these laws lack the details to address real-time deepfake issues.

\subsubsection{Raising Public Awareness.}
After a deepfake spreads, raising public awareness is crucial to minimize harm and prevent victim-blaming. Campaigns should educate the public about the potential for manipulation, highlighting that anyone can be targeted by deepfake technology, thus discouraging victim-blaming \cite{YadlinSegal2020}. Additionally, raising awareness about legal avenues for redress can empower victims, ensuring they know how to report and seek justice for harm caused by deepfakes \cite{Hoofnagle2019}. Promoting media literacy and fact-checking through trusted platforms like PolitiFact \cite{Masood2022} is essential for encouraging users to verify suspicious content. Platforms like Facebook collaborate with fact-checking organizations to label manipulated content and provide context, helping users identify deepfakes and understand the risks \cite{Tan2024}. Additionally, legal frameworks, such as the EU Digital Services Act\footnote{https://eur-lex.europa.eu/eli/reg/2022/2065/oj}, enable swift content removal and accountability for perpetrators, while blockchain-based content provenance systems ensure transparency and traceability \cite{Masood2022}. Platforms can track digital footprints and metadata to trace the origins of manipulated content, aiding in identifying creators or distributors \cite{Whyte2020}.  AI-driven tools like Sensity AI\footnote{https://sensity.ai/} and Deepware Scanner\footnote{https://scanner.deepware.ai/} help detect deepfakes in real-time, preventing further spread \cite{SamuelOkon2024}. These combined efforts protect victims and empower users to combat the impact of deepfakes.

\subsubsection{Content Monitoring and User Verification.}
Content monitoring and user verification can play a crucial role in identifying offenders and supporting legal cases after a deepfake has spread. Platforms can track the origins of deepfakes by analyzing the metadata and digital footprints left by the content's creator, helping identify the individual responsible for its creation or distribution \cite{Whyte2020}. User verification methods, such as requiring verified accounts for content upload, can deter malicious actors and provide a clear chain of accountability. When deepfakes are flagged, the platform's monitoring tools can cross-check the content against known sources and user reports to assess its authenticity, aiding law enforcement in identifying perpetrators. Additionally, ensuring the authenticity and integrity of media content involves advanced technologies like digital certificates, cryptographic hashes, and AI-based authentication tools. For example, Microsoft Azure uses digital certificates and cryptographic hashes to verify content authenticity \cite{SamuelOkon2024}, while AI tools like Microsoft Video Authenticator detect deepfake artifacts and provide confidence scores to alert users to potential manipulations \cite{Farish2019}. Platforms also integrate real-time content verification systems, flagging suspicious media upon upload to proactively prevent the spread of manipulated content. Blockchain-based systems, like Project Origin from the BBC, establish secure media provenance, offering irrefutable evidence of manipulation and origin, supporting legal claims and ensuring that victims have the necessary tools to seek justice \cite{Masood2022,KiChan2020}.

\subsubsection{Victim Support and Rehabilitation.}
After a deepfake spreads, providing comprehensive support and rehabilitation to victims is essential, as they often face significant emotional and psychological distress. Legislative frameworks must be updated to criminalize the malicious creation and distribution of deepfakes, offering victims clearer legal recourse \cite{RomeroMoreno2024}. Legal aid services are crucial to help victims navigate the complex legal landscape. Mental health support, including specialized counseling and therapy, is vital for recovery \cite{Mania2022}. Platforms should establish rapid response systems for removing harmful content, enabling victims to flag deepfakes for quick removal, while automated detection tools can proactively identify them. Collaboration among legal teams, tech companies, and mental health professionals can streamline this process. NGOs should work with governments, law enforcement, and tech firms to provide immediate assistance through hotlines, helping to remove victims' images from platforms without delay \cite{Haimson2021}, and offer educational resources on victims' rights and available support services \cite{Jiang2024}. These efforts ensure a holistic recovery approach addressing legal, psychological, and technical needs.

\section{Discussion}

Deepfakes, as revealed in this study, represent not only a technical threat but a systemic risk to the sociopolitical and informational stability of the Global South. The challenges are compounded in environments characterized by limited digital literacy, underdeveloped technological infrastructure, and fragile media ecosystems. The data shows a disturbing lack of individual confidence in deepfake detection: only a small percentage of respondents expressed high confidence in their ability to identify synthetic media, with 16.4\% reporting low confidence and 4.1\% stating no confidence at all. Additionally, 45.2\% of respondents reported only moderate confidence in their peers' detection skills. In many parts of the Global South, these figures are not just statistics but signals of systemic vulnerability. The inability to recognize manipulated content can amplify the risk of misinformation during critical periods such as elections, public health crises, or communal tensions. Integrating this insight into the "Prevention" layer of the framework, public awareness and education must be localized, using vernacular languages, culturally relevant examples, and low-tech delivery modes (e.g., community radio, posters, SMS-based alerts) to reach those outside digital hubs. Ethical AI development initiatives must also prioritize inclusive design, ensuring that future detection and mitigation tools are informed by real-world user constraints in low-resource settings.

Equally concerning are the gaps in information verification practices, which further compound the deepfake threat. While 69.9\%of survey participants claimed they cross-check content, a troubling 21.9\%admitted they do not verify information at all. This is particularly dangerous in societies where information verification is not institutionalized, and where traditional fact-checking infrastructure is either absent or inaccessible. In such settings, misinformation, amplified by deepfakes,, can spread rapidly via community networks, often being accepted as truth due to existing trust in peer-shared information. The framework's Detection pillar, which emphasizes accessible, AI-powered verification tools, becomes essential here. These tools must be embedded directly into platforms like WhatsApp, Telegram, and Facebook, which are heavily used in the Global South. Moreover, detection systems should be lightweight enough to function on low-bandwidth connections and affordable devices. Training initiatives should also target specific community segments, such as youth, teachers, and religious leaders, who often serve as information gatekeepers in these societies. These groups can act as force multipliers in disseminating detection practices and cultivating skepticism toward viral or AI-generated content.

Our survey highlights a critical erosion of trust in media systems, with 91.8\% of respondents concerned about deepfakes being used to deceive the public. In many Global South nations, where media freedom is constrained and disinformation is weaponized, deepfakes risk both manipulating opinion and accelerating democratic backsliding. While healthy skepticism has value, it becomes corrosive when audiences cannot distinguish real from fabricated content. The framework’s call for transparency via content labeling and public education is essential. Requiring outlets to disclose AI-generated or modified content can rebuild trust, supported by public service announcements, collaborations with local influencers, and media literacy in schools. Underfunded local media need help adopting authentication tools, and partnerships with global fact-checkers can strengthen capacity.
Public perceptions of AI detection tools present another challenge: while half support their use, skepticism persists due to limited exposure and low digital trust. Accessible design and user education are key, smartphone apps with confidence scores, embedded tutorials, and localized training hubs can aid adoption. Tools must be updated regularly to keep pace with deepfake advances, requiring collaboration among researchers, regulators, and civil society, with feedback loops enabling user participation.
On regulation, 63\% support stronger deepfake controls, though broader AI regulation draws mixed reactions due to innovation concerns. Frameworks should target malicious uses without penalizing satire, education, or cultural projects. A graduated oversight model, strict for political deepfakes, lenient for low-risk uses, can balance protection with innovation. Co-regulation and inclusive national AI councils can ensure policies reflect grassroots needs, with cross-border cooperation promoting harmonized standards.
Trust restoration strategies, supported by 68.5\% of respondents, must be continuous. Media outlets should adopt open practices like provenance trails, while governments issue timely corrections. Civil society can run forums and town halls on deepfake detection, and academia can study trust dynamics over time. The framework’s prevention-detection-mitigation model will work only if grounded in local realities, fostering an ecosystem resistant to deepfakes and resilient in upholding truth.

In urban areas with high digital media consumption, public education can help citizens identify and report deepfakes, while international collaboration is essential to address this global challenge. We also explore the ethical dilemmas surrounding deepfakes, particularly in regions with limited digital literacy, and propose a balanced approach to mitigate harms while recognizing the legitimate uses of AI.
We aim to inform policymakers, educators, and tech developers about the risks of deepfakes in developing countries, emphasizing the need for a collaborative effort to create effective solutions. A challenge in our research was the varying awareness levels among participants, with some recognizing the risks while others were unaware. Our urban focus also limited perspectives from rural areas, where digital literacy challenges and deepfake exposure may differ.


\section{Next Steps and Future Work}
Future research should expand beyond urban populations to examine rural and underserved communities in the Global South, where deepfakes may have distinct impacts due to limited infrastructure, lower literacy, and reduced access to verification tools. Mixed-methods approaches, including surveys, interviews, and ethnographic fieldwork, can capture the complex ways deepfakes intersect with local sociopolitical dynamics, especially among marginalized groups such as women, linguistic minorities, and the elderly. Collaborations with local NGOs and community media can ensure research instruments and interventions are culturally relevant. Longitudinal studies tracking awareness, attitudes, and behavioral metrics, such as misinformation-sharing patterns or response times to deepfake exposure, would help assess the evolving effectiveness of educational, policy, and technological measures. Technological innovation should prioritize lightweight, multilingual, mobile-compatible, and offline-capable detection tools, integrated into widely used applications like WhatsApp or Telegram, with explainable outputs to foster trust. Regional tech hubs and universities can contribute localized datasets and culturally adapted models to improve detection accuracy. Policy research should test co-regulatory frameworks involving governments, civil society, and technology developers, evaluating pilot enforcement and media labeling protocols for replication. Interdisciplinary collaboration between AI researchers, legal experts, educators, and community leaders will be essential to building resilient, inclusive responses that protect democratic integrity.

\section{Conclusion}
This research sheds light on the multifaceted risks posed by deepfake technology in developing regions of the Global South, where limited digital literacy and technology access exacerbate its harmful potential. By exploring the challenges faced in these areas, we underline the importance of targeted public education, robust legal frameworks, and international collaboration to mitigate the impact of deepfakes. Our findings highlight the need for proactive measures to safeguard political stability, social trust, and individual well-being. Moving forward, it is crucial to address the varying awareness levels among populations, particularly in rural contexts, where exposure to and understanding of digital threats may differ. We encourage future work to explore these disparities further, developing solutions tailored to diverse contexts within the Global South.



\bibliographystyle{ACM-Reference-Format}
\balance
\bibliography{our_work}

\appendix
\section{Questionnaire} \label{apx:Questionnaire}
Here is the full survey questionnaire:

\begin{enumerate}
    \item Age:
    \\ \textit{Options:} \begin{inparaenum}[i)]
        \item Under 18
        \item 18-24
        \item 25-34
        \item 35-44
        \item 45 and above
    \end{inparaenum}
    \item Gender:
    \\ \textit{Options:} \begin{inparaenum}[i)]
        \item Male
        \item Female
        \item Other (Please write your answer .........)
    \end{inparaenum}
    \item Highest Education Level:
    \\ \textit{Options:} \begin{inparaenum}[i)]
        \item No formal education
        \item Primary school
        \item Secondary school
        \item Higher education (undergraduate, postgraduate)
        \item Other (Please write your answer .........)
    \end{inparaenum}
    \item Occupation:
    \\ \textit{Options:} \begin{inparaenum}[i)]
        \item Student
        \item Corporate professional
        \item Educator
        \item Media/Journalism
        \item IT/Technology
        \item Other (Please write your answer .........)
    \end{inparaenum}
    \item What type of area do you currently live in?
    \\ \textit{Options:} \begin{inparaenum}[i)]
        \item Urban
        \item Suburban
        \item Rural
        \item Other (Please write your answer .........)
    \end{inparaenum}
    \item Country:
    \item Have you heard of the term \textbf{``deepfake''}? Do you know what it means?
    \\ \textit{Options:} \begin{inparaenum}[i)]
        \item Yes
        \item No
    \end{inparaenum}
    \item How would you describe your level of understanding of deepfakes?
    \\ \textit{Options:} \begin{inparaenum}[i)]
        \item Heard the term but not sure what it means properly
        \item Somewhat knowledgeable
        \item Very knowledgeable
    \end{inparaenum}
    \item Where have you learned about deepfakes?
    \\ \textit{Options:} \begin{inparaenum}[i)]
        \item Social media platforms (e.g., Facebook, Twitter, TikTok)
        \item News outlets (online, TV, or print)
        \item Educational resources (e.g., articles, courses, or lectures)
        \item Friends or family
        \item Online videos or tutorials (e.g., YouTube)
        \item Personal or professional experience
        \item I have not heard about deepfakes before this survey
    \end{inparaenum}
    \item Have you ever encountered a deepfake (in videos, audio, or images)?
    \\ \textit{Options:} \begin{inparaenum}[i)]
        \item Yes
        \item No
    \end{inparaenum}
    \item If yes, where did you encounter it?
    \\ \textit{Options:} \begin{inparaenum}[i)]
        \item Social media
        \item News media
        \item Video platforms
        \item Other (Please write your answer .........)
    \end{inparaenum}
    \item When you see a video or image online, how confident are you that it is authentic? (Not confident at all to Very confident)
    \\ \textit{Options:} \begin{inparaenum}[i)]
        \item 1
        \item 2
        \item 3
        \item 4
        \item 5
    \end{inparaenum}

    \item If yes, in which context did you encounter it? (Select all that apply)
    \\ \textit{Options:} \begin{inparaenum}[i)]
        \item Social media (e.g., fake videos/images of public figures)
        \item Entertainment (e.g., movies, memes, parodies)
        \item Politics (e.g., fake speeches or announcements)
        \item Cybersecurity (e.g., scams, phishing, identity theft)
        \item Personal experience
        \item Other (Please write your answer .........)
    \end{inparaenum}
    \item How much deepfake content do you think is available on social media?
    \\ \textit{Options:} \begin{inparaenum}[i)]
        \item Very little
        \item 1
        \item 2
        \item 3
        \item 4
        \item Almost all content
    \end{inparaenum}
    \item Did any deepfake content ever have a significant impact on you?
    \\ \textit{Options:} \begin{inparaenum}[i)]
        \item No, it was harmless
        \item Unsure
        \item Yes, it was harmful
    \end{inparaenum}
    \item If it was harmful, what type of harm did it cause?
    \\ \textit{Options:} \begin{inparaenum}[i)]
        \item No, it was harmless
        \item Yes, it was emotionally distressing
        \item Yes, it was physically distressing
        \item Yes, it caused financial loss or harm
        \item Other (Please write your answer .........)
    \end{inparaenum}
    \item Did any deepfake content ever have a significant impact on your friends or family?
    \\ \textit{Options:} \begin{inparaenum}[i)]
        \item No, it was harmless
        \item Unsure
        \item Yes, it was harmful
    \end{inparaenum}
    \item How harmful do you think deepfakes can be? (Very harmful to Not harmful at all)
    \\ \textit{Options:} \begin{inparaenum}[i)]
        \item 1
        \item 2
        \item 3
        \item 4
        \item 5
    \end{inparaenum}
    \item Do you think deepfakes could be used to mislead people in your community?
    \\ \textit{Options:} \begin{inparaenum}[i)]
        \item Yes
        \item No
        \item Not sure
    \end{inparaenum}
    \item How likely do you think deepfakes are to influence public opinion or decisions in your community? (Very likely to Not likely at all)
    \\ \textit{Options:} \begin{inparaenum}[i)]
        \item 1
        \item 2
        \item 3
        \item 4
        \item 5
    \end{inparaenum}
    \item How concerned are you about the misuse of deepfakes?  (Not concerned at all to Very concerned)
    \\ \textit{Options:} \begin{inparaenum}[i)]
        \item 1
        \item 2
        \item 3
        \item 4
        \item 5
    \end{inparaenum}

    \item Do you trust information from the following sources? (Rate each between: Very trustworthy, Somewhat trustworthy, Not trustworthy)
    \\ \textit{Options:} \begin{inparaenum}[i)]
        \item News outlets (TV, online news)
        \item Social media platforms (Facebook, Twitter, etc.)
        \item Teachers/Seniors
        \item Family/friends/neighbours
    \end{inparaenum}
    
    \item How do you typically verify the information you receive online? (Select all that apply)
    \\ \textit{Options:} \begin{inparaenum}[i)]
        \item I don't verify the information
        \item Cross-check with other sources
        \item Ask others if they think it's real
        \item Other (Please write your answer .........)
    \end{inparaenum}
    
    \item How confident are you in your ability to detect a deepfake? (Not confident at all to Very confident)
    \\ \textit{Options:} \begin{inparaenum}[i)]
        \item 1
        \item 2
        \item 3
        \item 4
        \item 5
    \end{inparaenum}
    
    \item How confident are you in your friends' or family's ability to detect a deepfake? (Not confident at all to Very confident)
    \\ \textit{Options:} \begin{inparaenum}[i)]
        \item 1
        \item 2
        \item 3
        \item 4
        \item 5
    \end{inparaenum}
    
    \item What do you think is the most effective way to combat deepfake-related issues? (Select all that apply)
    \\ \textit{Options:} \begin{inparaenum}[i)]
        \item Hardening AI Use Laws and Regulations
        \item Improved AI detection tools
        \item Stricter laws and regulations
        \item Public awareness campaigns
        \item Individual responsibility (e.g., verifying content before sharing)
        \item Other (Please write your answer .........)
    \end{inparaenum}
    
    \item What strategies do you think would help protect people from the negative impacts of deepfakes? (Select all that apply)
    \\ \textit{Options:} \begin{inparaenum}[i)]
        \item Social Campaigns (Education on identifying deepfakes)
        \item Improved regulations on social media
        \item More reliable fact-checking systems
        \item Improving rules and regulation regarding AI/AI Tools
        \item Other (Please write your answer .........)
    \end{inparaenum}
    
    \item Do you believe your workplace or educational institution provides sufficient training or awareness about deepfakes?
    \\ \textit{Options:} \begin{inparaenum}[i)]
        \item Yes
        \item No
        \item Unsure
    \end{inparaenum}
    
    \item What would you recommend improving trust in the media in your community? (Select all that apply)
    \\ \textit{Options:} \begin{inparaenum}[i)]
        \item Greater transparency from media sources
        \item Public awareness campaigns
        \item Technology to identify and flag deepfakes
        \item Stricter laws and regulations
        \item Other (Please write your answer .........)
    \end{inparaenum}
    
    \item How do you think AI can combat deepfakes or misinformation? (Select all that apply)
    \\ \textit{Options:}\begin{itemize}
        \item I don't believe AI can help combat deepfakes or misinformation
        \item Developing deepfake detection tools to identify manipulated content
        \item Creating AI systems that flag suspicious content in real-time
        \item Educating the public through AI-powered awareness campaigns
        \item Enhancing media verification and fact-checking with AI
        \item Using AI to track the origin and spread of misinformation
        \item Collaborating with social media platforms to remove deepfake content
        \item Implementing AI-based authentication systems for media
        \item Other (Please write your answer .........)
    \end{itemize}
\end{enumerate}

\end{document}